# Enhanced High-Field Transport Critical Current Densities Observed for the ex-situ PIT Processed Ag/ (Ba,K)Fe$_2$As$_2$ Thin Tapes

**Kazumasa Togano[1], Zhaoshun Gao[1], Hideaki Taira[2], Shigeyuki Ishida[2], Kunihiro Kihou[2], Akira Iyo[2] and Hiroshi Eisaki[2], Akiyoshi Matsumoto[1] and Hiroaki Kumakura[1]**

[1]National Institute for Materials Science (NIMS), Tsukuba Ibaraki 305-0047, Japan
[2]National Institute of Advanced Industrial Science and Technology (AIST), Tsukuba Ibaraki 305-8568, Japan

E-mail: TOGANO.Kazumasa@nims.go.jp

**Abstract**

We found that the transport $J_c$ of the ex-situ PIT processed (Ba,K)Fe$_2$As$_2$ (Ba-122) wire with single Ag sheath can be significantly enhanced by repeating the combined process of rolling and heat treatment. The transport $J_c$ (4.2 K and 10 T) of 4.4 x 10$^3$ A/cm$^2$ ($I_c$ =15.7 A) was obtained for a thin tape of 0.3 mm in thickness, which is the highest value reported so far for the PIT processed 122 wires and tapes with a Ag sheath and processed by a conventional route. The measurement by a hybrid magnet showed that $J_c$-$H$ curve maintains very small field dependence up to the strong magnetic field of 28 T as expected from the previously reported high $H_{c2}$ value. The core of the thin tape shows dense grain structure with less cracks and voids and indicates the development of c-axis alignment, although it is still incomplete. The researches to elucidate the origin of $J_c$ enhancement and to have further improvement of transport $J_c$ are now ongoing. The process is simple using a Ag single sheath and, therefore, more realistic technique for long length wire production.

## 1. Introduction

The discovery of superconductivity at 26 K in $LaFeAsO_{1-x}F_x$ [1] in early 2008 aroused enormous research on the iron-based superconductors and, to date, many families such as REOFeAs(1111-type)[2], LiFeAs(111-type)[3], $BaFe_2As_2$(122-type)[4], FeSe(11-type)[5] and the pnictides with perovskite-type blocking layer[6] have been found to show high temperature superconductivity up to 56 K by appropriate doping. In addition to the high transition temperature, $T_c$, the iron based superconductors were reported to have a very high upper critical field, $H_{c2}$, bringing the hope of high field applications as wires and bulks [7-9]. In order to evaluate the potentiality for wire applications, the development of wire processing technique is essential. Until now, powder-in-tube (PIT) process is most widely employed for the fabrication of 1111[10, 11], 11 [12] and 122 type [13-15] superconducting wires and tapes using various sheath materials such as Fe, Ag, Nb and Ta. However, the transport critical current densities, $J_c$, reported at an early stage were disappointing low due to the weak link grain boundary problem. This is arisen from the presence of numerous cracks and voids along the grain boundaries as well as the misorientation between the grains.

Very recently, there were some progresses in improving the grain connectivity of the PIT processed 122 wires. Weiss et al [16] reported the transport $J_c$ of as high as $\sim 8 \times 10^3$ $A/cm^2$ at 4.2 K and 10 T for the ex-situ PIT processed Cu/Ag/Ba-122 round wires, which was achieved by using a high quality precursor prepared by high energy ball milling and applying cold isotropic press (CIP) and hot isotropic press (HIP). The results indicate the importance of the quality of precursor and the densification of 122 core. On the other hand, Gao et al [17,18] demonstrated that the texturing is very effective to improve the grain connectivity as well as the densification. They obtained strong c-axis texturing for the ex-situ PIT processed Sr-122 with Sn addition. The sheath material used was Fe because the heat treatment was carried out at a high temperature for short time. They reported the $J_c$ value (at 4.2 K and 10 T) of $3.5 \times 10^3$ $A/cm^2$ [17] and more recently above $10^4$ $A/cm^2$ (at 4.2 K and 10 T) [18] which is the highest $J_c$ reported so far for the PIT processed iron based superconducting wires. Matsumoto et al [19] carried out the first measurement of transport $J_c$ at the temperatures above 4.2 K using a Ag/Ba-122 wire and found that the $J_c$-$H$ curve maintains small magnetic field dependence at temperatures up to 20 K. Those results indicate that the PIT processed 122 superconducting wires are promising for magnetic field applications at the medium temperature of cryogenic cooling or liquid hydrogen as well as in liquid helium.

For practical applications, the process should be simple and easy to be scaled up to the long length wire production, requiring little sophisticated equipments and processes. Like Bi-based high temperature superconductors, Ag looks the most suitable sheath material for 122 phase such as $(Ba,K)Fe_2As_2$ and $(Ba,Sr)Fe_2As_2$. This is because Ag does not react with 122 superconductors and is expected to serve as a stabilizing material of the conductor due to its low electrical resistivity. Very recently, Yao et al [20] reported the successful fabrication of multi-filamentary Sr-122 wires with an Fe/Ag double sheath.

In this paper, we demonstrate that a large enhancement of transport $J_c$ can be obtained for the ex-situ PIT processed Ag/Ba-122 tape by repeating the combined process of ordinal cold rolling and heat treatment. The method is basically the same with the commercial production process of $Bi_2Sr_2Ca_2Cu_3O_x$ (Bi-2223) superconducting tapes. We also report the critical current measurement in very high magnetic fields up to 28 T using a hybrid magnet in order to evaluate the high field performance of the PIT processed Ba-122 superconductor.

## 2. Experimental

Starting materials for the preparation of precursor were BaAs, KAs and $Fe_2As$ compounds, which were synthesized by reacting constituent elements. By using these compounds instead of each element, the loss of volatize elements such as K and As can be minimized. The powders of BaAs, KAs and $Fe_2As$ were mixed with the nominal composition of $(Ba_{0.6}K_{0.42})Fe_2As_{2.02}$ and reacted in an alumina crucible sealed in a stainless steel tube in a nitrogen atmosphere [21]. A small amount of excess KAs was added in order to compensate the loss of K and As through the reaction process. The assemble was then heat treated at 800ºC for 20 h to form Ba-122 phase. The obtained $(Ba_{0.6}K_{0.4})Fe_2As_2$ precursor has a fairly good quality as shown in the magnetization vs. temperature curve and the x-ray diffraction pattern of **Fig. 1**. The detail of the preparation will be published elsewhere.

The precursor was ground into a powder and mixed with Sn powder with the molar ratio of $(Ba_{0.6}K_{0.4})Fe_2As_2Sn_{0.5}$ using an agate mortar in a glove box filled with a high purity argon gas. The Sn was added in order to improve the grain connectivity, as reported for the Sr-122 system [17]. The powder mixture was packed into a Ag tube (outside diameter: 6 mm, inside diameter: 4 mm), which was subsequently groove rolled and swaged into a wire of 2 mm in diameter. A short piece of ~40 mm in length cut from the wire was then subjected to the first heat treatment of 850 ºC for 10 h for sintering. After the first heat treatment , the wire was deformed into a tape form using a flat rolling machine initially into 0.6 mm in thickness followed by the second heat treatment of 850 ºC for 10 h and finally into 0.3 mm in thickness followed by the third heat treatment of 850 ºC for 10 h. **Figure 2** shows the optical micrographs of the transverse cross sections observed for a round wire of 2mm in diameter, and the tapes of 0.6 mm and 0.3 mm in thickness, hereinafter abbreviated as 2 $mm^d$ wire, 0.6 $mm^t$ tape and 0.3 $mm^t$ tape, respectively. The photographs of the tapes show that the composite was deformed uniformly even after the previous heat treatment for sintering, similarly to the PIT processed Bi-2223 tape.

After each heat treatment, the $I_c$ measurement was carried out in liquid helium (4.2 K) using a 12 T superconducting magnet. The magnetic field was applied perpendicularly to the sample length and, in case of the tapes, parallel to the tape surface. $I_c$ was determined using the voltage criterion of 1 μm/cm. Transport critical current density, $J_c$, was estimated by dividing the $I_c$ by the cross sectional area of the superconducting core, which was measured by using the image analysis of a laser optical microscope. We also carried out the $I_c$ measurement in a 28 T hybrid magnet of the Tsukuba Magnet Laboratory (TML) of National Institute of Materials Science (NIMS). The cross sections of the wire and tapes were observed by an optical microscope (OM) and a scanning electron microscope (SEM). The phase identification was carried out by X-ray diffraction (XRD) analysis.

## 3. Results and Discussion

In the transport $I_c$ measurement, the voltage versus applied current curves show rather sharp transition as shown in **Fig. 3 (a)**. **Figure 3 (b)** is the plots of transport $J_c$ as a function of magnetic field of the 2 mm$^d$ round wire, and the 0.6 mm$^t$ and 0.3 mm$^t$ tapes. The $J_c$ of the 2 mm$^d$ round wire in self field is $7.4 \times 10^3$ A/cm$^2$ ($I_c$ = 67.5 A). However, in applied magnetic fields, it rapidly drops more than one order of magnitude and then shows very small field dependence in the strong magnetic field region. The $J_c$ of the 2mm$^d$ round wire at 10 T is as low as $7.6 \times 10^2$ A/cm$^2$ ($I_c$ = 5.5 A), which is comparable to those reported so far for Ag sheathed PIT 122 round wires processed by the conventional PIT route [14,15]. However, figure 3 (b) shows that the transport $J_c$ can be drastically increased after the rolling and subsequent heat treatment. The enhancement is significant in applied fields, while the $J_c$ in self filed is not so influenced by the rolling and annealing. The $J_c$ value at 10 T is increased by about three times ($2.1 \times 10^3$ A/cm$^2$ ($I_c$ = 10.6 A)) for the 0.6 mm$^t$ tape and further about two times ($4.4 \times 10^3$ A/cm$^2$ ($I_c$ = 15.7 A)) for the 0.3 mm$^t$ tapes. The $J_c$ of $4.4 \times 10^3$ A/cm$^2$ (10 T and 4.2 K) is the highest value reported so far for the Ag-sheathed wires and tapes processed by the conventional PIT route without high pressure techniques. For the practical application of tape conductors, one of the concerns is the anisotropy of $I_c$. We measured the $I_c$ of the 0.3 mm$^t$ tape in the field applied perpendicularly to the tape surface and found that the $I_c$ becomes smaller in the entire magnetic fields than that in the parallel field. The ratio of $I_c$(perpendicular)/$I_c$(parallel) is 0.5 at 10 T. This is contrary to the results for the Fe-sheathed tape in which the $J_c$ is higher in the perpendicular fields [18]. The difference is probably caused by the different processing conditions

In order to check the effectiveness of the intermediate rolling and annealing, we prepared a reference sample without intermediate step, that is, by rolling the 2 mm$^d$ wire directly to a 0.3 mm$^t$ tape followed by annealing at 850 °C for 10 h. The tape shows the transport $J_c$ values less than $10^3$ A/cm$^2$ at 4.2 K and 10 T with poor reproducibility of the result. This is similar to the Bi-2223 tape fabrication, in which too large reduction ratio before the heat treatment results in an inhomogeneous deformation and poor $J_c$ and, hence, the tape fabrication is carried out by repeating the intermediate rolling with smaller reduction rate and annealing [22]. Optimization of various processing parameters for Ag/Ba-122 tape such as reduction ratio and heat treatment condition is a future research issue.

$I_c$ measurement using a 28 T hybrid magnet was carried out at 4.2 K for the 0.3 mm$^t$ tape sample used for the measurement of a 12 T superconducting magnet. **Figure 4** shows the result together with the data measured with the 12 T superconducting magnet. Typical $J_c$-$H$ curves of the bronze processed (Nb,Ti)$_3$Sn [23] and PIT processed MgB$_2$ [24] superconducting wires are also included for comparison. The data of the hybrid magnet and those of the 12 T superconducting magnet coincide with each other indicating that no degradation occurred during the duration between both measurements (~1 month). It is noteworthy that the $J_c$-$H$ curve shows extremely small magnetic field dependence up to the strong magnetic field of 28 T. The $J_c$ at 28 T is 2.9x10$^3$ A/cm$^2$ ($I_c$ = 10.5 A), which is still two-thirds of that at 10 T. The $J_c$-$H$ curve crosses those of (Nb,Ti)$_3$Sn and MgB$_2$ wires at magnetic fields far below 28 T. This is consistent with the very high $H_{c2}$ of 122 phase as reported in earlier papers [7-9] and shows the good potentiality of 122 wires for high field generation.

In order to investigate the mechanism of $J_c$ enhancement, we carried out the microstructure observations. **Figure 5** shows the comparison of high magnification optical micrograph between the 2mm$^d$ wire and the 0.3 mm$^t$ tape. Both structures are composed of the matrix of 122 grains and the dispersed particles which appear white in the figures. However, we can see some differences when looking into the details. One is that the cracks along the grain boundary, which are observed frequently in the 2mm$^d$ wire as indicated by arrows in the Fig. 5 (a), are rarely observed in the 0.3$^t$ mm tape. The other is the grain structure of 122 phase. The grains of the 2mm$^d$ wire are equi-axed and the average grain size is ~5.6 μm. On the other hand, the 0.3 mm$^t$ tape shows more uniform structure with slightly larger grain size. Furthermore, the average grain size shows a slight difference when measured in the directions parallel and perpendicular to the rolling plane, which are 7.5 μm and 5.8 μm, respectively. This indicates that the grains are slightly elongated by the rolling process, although it is not so significant as observed in the Fe-sheathed Sr-122 tapes [17] and Bi$_2$Sr$_2$Ca$_2$Cu$_3$O$_x$ tapes [22].

**Figure 6** shows the SEM and EDAX mapping images of the constituent element observed on the cross section of the 0.3 mm$^t$ tape. The result of the mapping shows that the white particles observed in the OM and SEM micrographs are Ag based alloy. This is surprising because no Ag was added to the starting powder mixture. The point analysis on a white particle indicates that the composition is Ag-7.5%Sn-2.7%As-2.5%Fe.  It is not clear at present how these Ag particles are included in the core.   We suppose that Ag-Sn alloy liquid was formed by the reaction between the added Sn and the Ag sheath infiltrating inside the core and precipitated as the observed particles.  The dissolution of 7%Sn to Ag leads to ~100°C drop of the melting point.   The role of Ag-Sn alloy for the improvement of grain connectivity is interesting future subject to be investigated.

Anisotropic nature of the tape was also observed by the x-ray diffraction.  **Figure 7 (a) and (b)** show the XRD patterns of the 0.3 mm$^t$ and 0.6 mm$^t$ tapes taken on the core surface which was exposed when the sheath was mechanically separated.   XRD pattern of the starting powder mixture of $(Ba,K)Fe_2As_2$+Sn was also shown in **Fig. 7 (c)** for comparison.   Relative intensities of all basal planes (00l) of the tapes are apparently stronger than those of randomly oriented powder mixture.  This indicates that the compressive stress of rolling deformation forces the basal plane of 122 grain to align parallel to the rolling plane.   The texture obtained in this study is still incomplete compared to that achieved in the Bi-2223 superconducting tape, for which the repetition of the rolling and annealing is essential to have strong texturing and high performance of critical current property [25].  The peaks of Sn in the starting powder mixture are not observed in the tapes and, instead, Ag-based alloy peaks appear and become stronger by repeating the heat treatment.   This is in accordance with the precipitation of Ag-Sn alloy in the tapes observed by OM and SEM-EDAX.

From the microstructure investigations, two factors emerge as the possible origin of $J_c$ enhancement.   One is the reduction of cracks and voids and the other is the texture of c-axis grain alignment.   The effectiveness of the texturing was already demonstrated by the work for Fe-sheathed Sr-122 tape with Sn addition, in which the strong texture was achieved solely by applying rolling process [17, 18].   From this result, we can expect to have much higher transport $J_c$ for Ag-sheathed 122 wires by achieving higher degree of c-axis texturing.   Since Ag and Fe have different crystal structures, which have influence on the various mechanical behaviors such as hardness and work hardening, optimization of processing parameters to have stronger texture must be carried out independently of each other.   We believe that our results give a progress in the development of the PIT processed Ag-sheathed 122 tapes with improved high field characteristics.

## 4. Conclusions

We have demonstrated that the transport $J_c$ of the Ag single sheathed $(Ba,K)Fe_2As_2$ tape can be significantly enhanced by repeating the combined process of rolling and heat treatment.   The transport $J_c$ (4.2 K and 10 T) of 4.4 x $10^3$ A/cm$^2$   ($I_c$ =15.7 A) was obtained for a thin tape of 0.3 mm in thickness, which is the highest value reported so far for the Ag sheathed 122 wires and tapes processed by conventional PIT route in normal pressure.   We also found that $J_c$-H curve shows very small field dependence up to the strong magnetic field of 28 T as expected from the previously reported high $H_{c2}$ value. The result of microstructure observation shows that the core of the tape has much less cracks and voids and the indication of the texture development by rolling. The texture is still incomplete compared to those of Bi-based superconductors and, therefore, we expect to have much higher transport $J_c$ by achieving higher degree of c-axis texturing. We believe that the process presented in this paper is simple using the Ag single sheath and, therefore, more realistic for long length wire production.

**Acknowledgements**

This work was supported the Japan Society for the Promotion of Science (JSPS) through its "Funding Program for World-Leading Innovative R & D on Science and Technology (FIRST) Program". We acknowledge Dr. H. Fujii and Mr. S.J. Ye of National Institute for Materials Science, Tsukuba, Japan for their assistance in $I_c$ measurement and SEM-EDAX observation

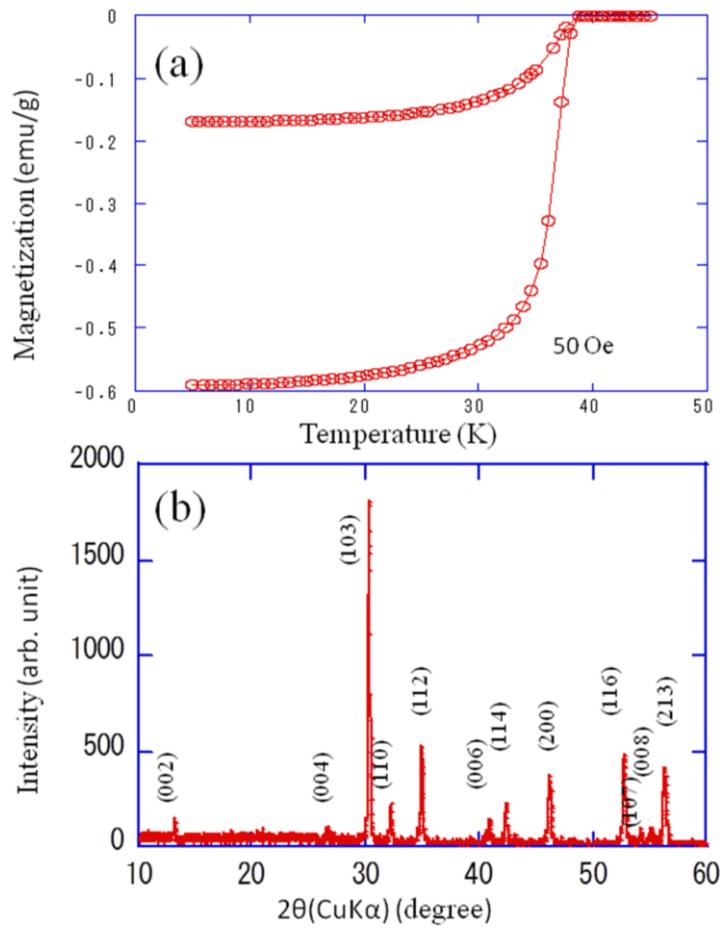

**Figure 1**  (a) Magnetization versus temperature curve, and (b) x-ray diffraction pattern of $(Ba_{0.6}K_{0.4})Fe_2As_2$ precursor.

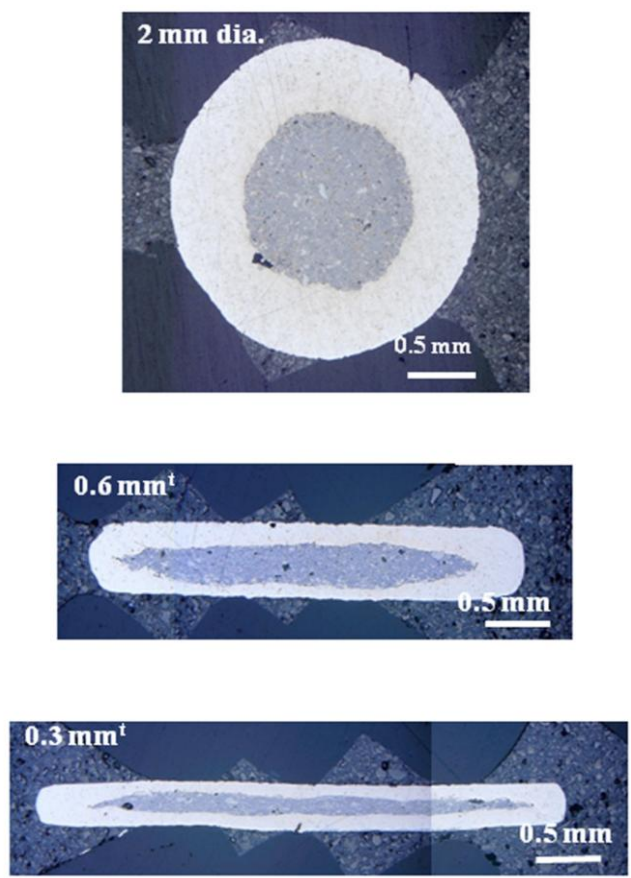

**Figure 2** Cross sections of the wire of 2 mm in diameter and the tapes of 0.6 and 0.3 mm in thickness.

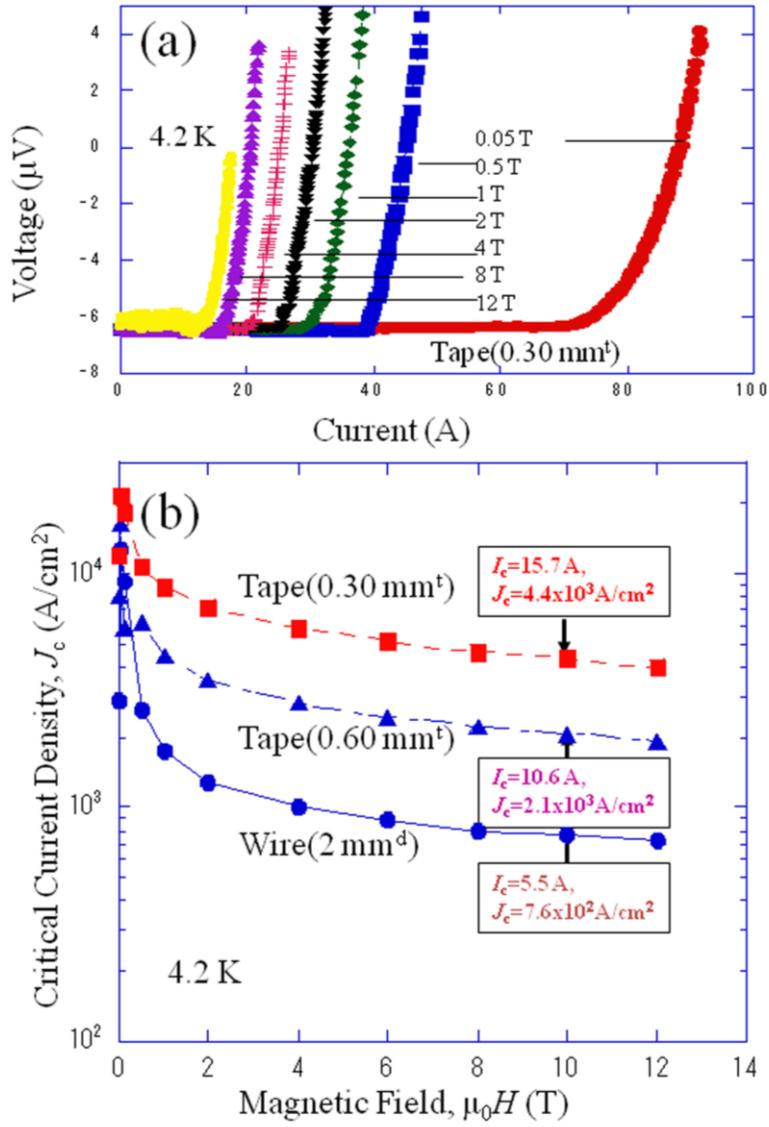

**Figure 3** (a) Voltage versus current curves for the 0.3 mm$^t$ tape, (b) plots of transport critical current density $J_c$ as a function of applied magnetic field for the 2mm$^d$ round wire and the 0.6 and 0.3 mm$^t$ tapes.   The measurement was carried out using a 12 T superconducting magnet.

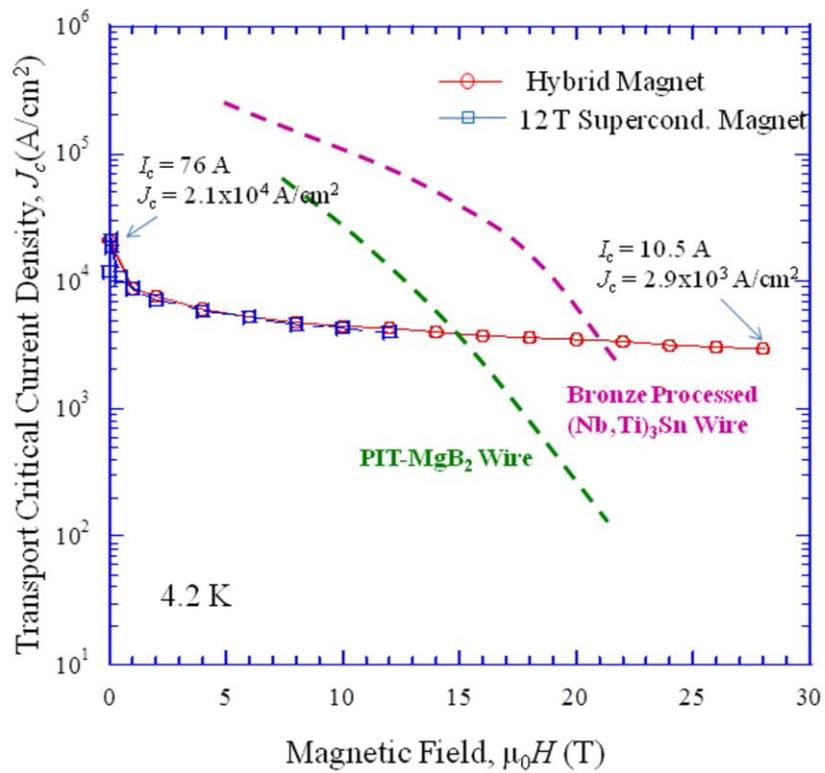

**Figure 4** Transport $J_c$ versus applied field curve of the 0.3 mm$^t$ tape measured in a 28 T hybrid magnet. The data measured in a 12 T superconducting magnet and typical curves for bronze processed $(Nb,Ti)_3Sn$ [23] and PIT-$MgB_2$ [24] superconducting wires are also shown for comparison.

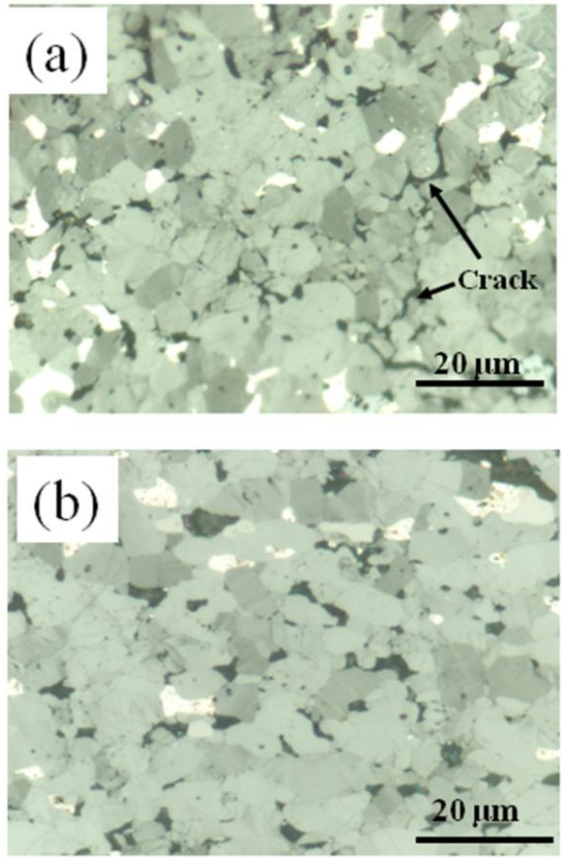

**Figure 5** Optical micrographs observed on the polished cross section of (a) the 2 mm$^d$ wire and (b) the 0.3 mm$^t$ tape.

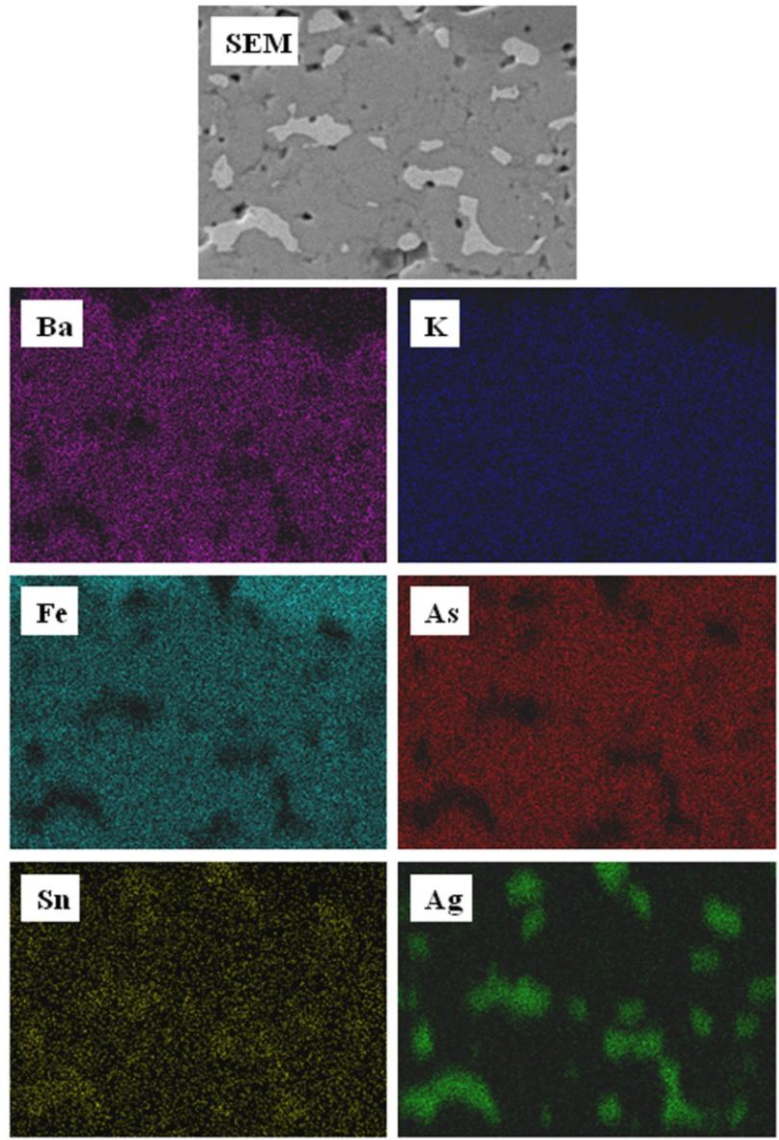

**Figure 6** Micrograph of SEM and mappings of constituent elements detected by a SEM-EDAX on the polished cross section of the 0.3 mm$^t$ tape.

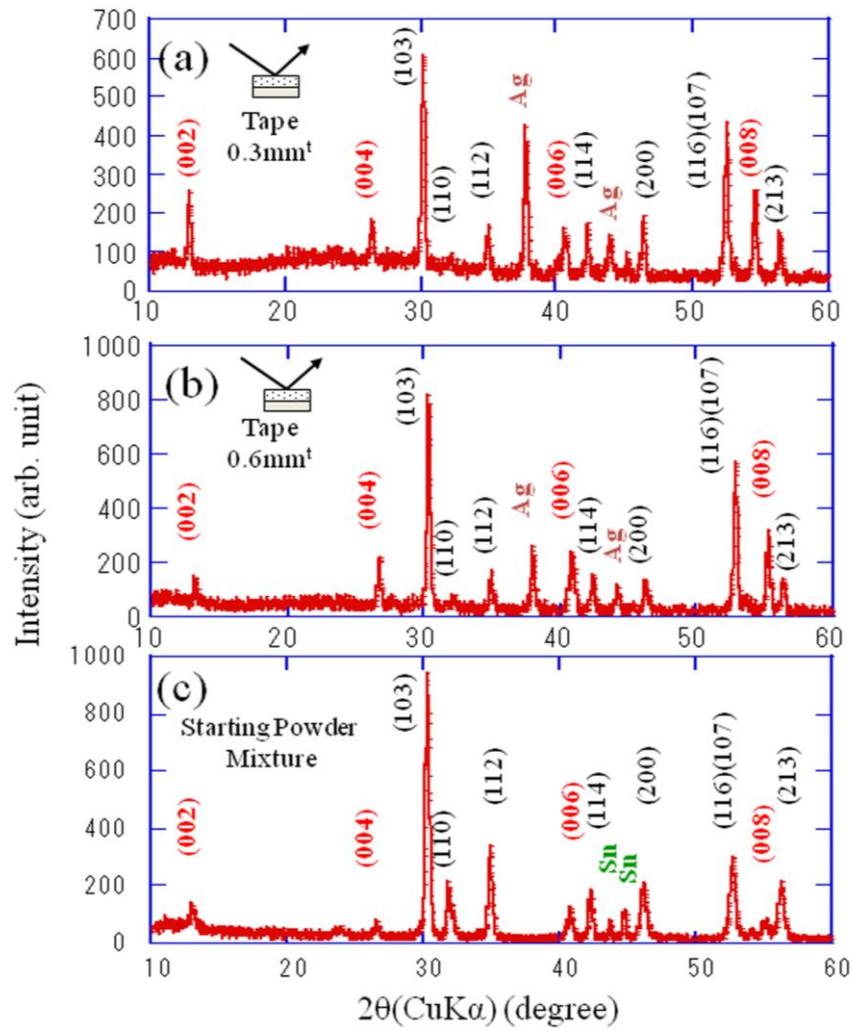

**Figure 7**  X-ray diffraction patterns for (a) the 0.3 mm$^t$ and (b) 0.6 mm$^t$ tapes.   The measurement was done on a surface of the core exposed after the sheath was mechanically separated. The pattern of a starting powder mixture of (Ba,K)Fe$_2$As$_2$+Sn is also shown in (c) for comparison